\documentclass[dvips]{article} 
\usepackage{spie}   
\usepackage{psfig}   
\def\REF{}
\def\fig{}

\title{Topic maps for custom viewing of data} 

\author{Ashish Mahabal\supit{a}, S. G. Djorgovski\supit{a},
Roy Williams\supit{a} and Robert Brunner\supit{a} 
\skiplinehalf 
\supit{a}Caltech, 1200 E California Blvd., Pasadena, CA, 91125, USA
}

\authorinfo{Contact information: Send correspondence to Ashish Mahabal.
E-mail: aam@astro.caltech.edu}

\begin{document} 
  \maketitle 

\begin{abstract}


A Topic Map is a structured network of hyperlinks that points into an information pool.
Topic Maps have an existence independent of the information pool and hence
different Topic Maps can form different layers above the same information pool
and provide us with different views of it. We explore the use of Topic Maps
with the Unified Column Descriptor (UCD) scheme developed in the frame of the
ESO-CDS data mining project. UCD, with its multi-tier hierarchical structure,
categorizes parameters reported in tables and catalogs. By using Topic Maps
we show how columns from different catalogs with similar but not identical
descriptions could be combined. A direct application for the Virtual
Observatory community is that of merging catalogs in order to generate
customized views of data.

\end{abstract}


\keywords{Virtual Observatory, Topic Map, XML, Large Datasets,
Knowledge Organization, Data Discovery, Semantic Nets}

\section{INTRODUCTION}
\label{sect:intro}  

Over the last couple of years several VO groups have been making
steady progress in defining and prototyping various access
and interconnectivity tools. It is clear that not only will the VO be
internet-based, but also that it will be very XML-heavy. 
The most important feature of the VO will be to allow access to desired
information in a quick and systematic way.
All the 
metadata - handle to the real data - will populate different XML files.
XML is highly structured and very readable. However, it does not offer
itself to searches trivially. A Topic Map (TM) - an XML file itself - can
form a layer of hyperlinks over an information resource and serve as 
the GPS in a huge database.
With TMs one can define complex knowledge
structures and attribute them as metadata to information resources
allowing one to systematically organize knowledge on a variety of data subjects
such that the retrieval and sharing with other users is easy.

As an analogy from everyday life, a Topic Map is like the index at the back
of a book. Like an index, a Topic Map presents an array of subjects, along
with links to the occurrences of information, which for a book are page
numbers. A book index can have entries in italic (indicating, for example,
the presence of an image on page so and so),
or in bold (the main explanation), and links to other parts of the index
(for `Active Galaxies', see `Galaxies, Active'). But Topic Maps also
provide different kinds of indexes for different purposes, for different
users. A TM links all the topics in a given area, letting users find the
information they want more transparently by navigating through the concepts
of the subject area, rather than having to know the underlying way in which
the information has been organized and stored.

Using Topic Maps to interlink datasets goes one step ahead than existing
tools and services for two reasons: (1) users will be able to customize
Topic Maps to the level that they desire, and (2) new VO services that
are developed elsewhere and by other teams will not only be instantly
linkable through Topic Maps, but also the Topic Map indexing and querying
mechanisms will allow a seamlessly merged appearance of these services
allowing for more semantically coherent knowledge discovery.

In this paper we provide an outline of what Topic Maps are and where 
the Topic Map technology is heading. We then describe the Uniform Column
Descriptors (UCDs) and go on to discuss the Topic Map made from the UCDs.
We then describe the Topic Map engine we have implemented to
allow users to create their own UCD Topic Maps and merge those with the ones we
have constructed.
Finally we present sketches for more Topic Maps which can be used as VO tools.

\section{Topic Maps}
\label{sect:topicmaps}  

Topic Maps started life as a standard for software documentation 
in 1991 (HyTime Hypertext 1991 conference in San Antonio). However,
that was much before the time was ripe for the idea. It was only when
XML matured that their potential in information and knowledge
organization has been realized. In 1996 Topic Maps
became a work item in ISO's SGML working group resulting in a Topic Map
standard published in 1999. This has since then led to different working
groups on XTM, TMQL and so on and the progress has been rapid.
\REF More history can be found in Pepper(1999). \REF

A {\it Topic Map} is a collection of {\it topics} linked together by
{\it associations}
between the topics. The topics can occur in different contexts ({\it scopes})
and the associations qualify the occurrences.
In the map, each node is a named topic.
Associations between nodes are expressed
by links. 
A Topic Map thus 
is a structured network of hyperlinks above an information
pool from which the topics are drawn. 
The named topics can be just about anything thereby
allowing a Topic Map to discuss abstract relationships between different
topics.
Depending on the situation, different terms could be chosen to be
regarded as topics. In the following sections we will describe in detail
the UCD Topic Map. For a more detailed, general introduction to Topic Maps,
please see \REF Mahabal et al.(2001).

The real power of Topic Maps comes from the scoping ability. Scopes can be used
to categorize different occurrences of a given topic depending on context. 
A good way to look at Topic Maps with their scoping ability is the following analogy:
When searching for something on the internet search engines, one gets a lot
of muck in which the real gems are hidden. If proper keywords are used, and if the
search engine does not index everything, the gems float to the top. This is exactly 
what happens when the scoping mechanism is used for discrimination and for
qualifying context. It limits the validity of a topic by categorizing it into one
or more themes. Thus scoping
allows one to zoom in not only on different topics but also very specific
subcategories e.g. a quasar could be scoped to be one with a binary companion,
and also one with a redshift exceeding 5. It becomes very useful when, for
example,  searching for specialized entities like high redshift quasar clusters.

In general, a Topic Map can be looked at in many different ways:
\begin{itemize}
\item As a semantic network over an information pool. Though a Topic Map 
connects (associates) different concepts/subjects (topics) within an
information pool, it has an existence distinct from the information
pool itself. This network of hyperlinks can be used in its own right as
a resource.
\item A configurable data interconnections viewer. Besides being a network of
hyperlinks, the Topic Map is also capable of rendering itself as a
configurable viewer of the data resource it is talking about. This comes
about because of the possibility of scoping which allows one to prune the 
Topic Map for viewing purposes in well defined complex ways.
\item Basis for structured querying of XML files.
This aspect of Topic Maps can not be overstated. Topic Maps
are wonderful for structuring and displaying information. In Sec.~6 we describe
Topic Map querying mechanisms.
\item A data/knowledge discovery tool. This is the aspect we want to
stress here. With the Virtual Observatory wanting to deal with extremely large datasets,
we need tools that will help us make sense of the multidimensionality (table-widths) as well
as the multi-trillion sizes of individual datasets (table-lengths). And then
we would want to combine the datasets meaningfully. What better way than to
do that while maintaining the semantic structure? Topic Maps allow for the diverse
data to be meaningfully combined.
\end{itemize}

The mathematical model that can be used for Topic Maps is that of a hypergraph.
Topics form the {\it vertices}, associations are the {\it edges} and 
occurrences form the {\it incidences}. A hypergraph representation
consisting of neighborhood of a vertex can be found at Kartoo\footnote{http://www.kartoo.com}.
An alternate representation of a hypergraph, a bipatriate graph, can
be seen as part of Mondeca's topic map software\footnote{http://www.mondeca.com}.
More details can be found in
\REF \REF Auillans et al.(2002) and
Auillans(2002).

\section{Unified Column Descriptors}
\label{sect:ucd}  

VO is about data: access to data, pipelines for data, processing of data,
interoperability and so on. The data itself is the main component and is
in the form of images, spectra, catalogs etc. An important component of all
these is the metadata. UCDs are a first step in bringing about uniformity
in the catalog metadata - the different column names that authors use. ESO
and CDS have worked together to form a 4-tier hierarchical set of standardized
column names - called Unified Column Descriptors - to categorize all the
different column names that are used in astronomical catalogs. There are 35 main
categories including Photometry, Positions, Physical Quantities etc. We have
made use of the 1400+ UCDs as our Published Subject Indicator (PSI) list in 
generating a template for the Topic Map we present here. Should the PSI evolve
to a different set of names, regenerating the Topic Map to reflect that is
straightforward. More details about
UCDs can be accessed at the Vizier website\footnote{http://vizier.u-strasbg.fr/doc/UCD.htx}, Ortiz et al.
(1999) \REF and \REF elsewhere in these proceedings
(Genova et al., 2002).
Some examples of UCDs are:
\begin{itemize}
\item PHOT\_FLUX\_IR\_12: This UCD is used for columns which represent 
``Flux density (IRAS) at 12 microns, or around 12 microns (ISO at 14.3)''. This pulls
together different names that different astronomers have used for this quantity e.g.
C, F12, F12umEst, FLUX12, Fnu\_12 etc. (see Fig.~3).
\item POS\_EQ\_RA\_MAIN: This UCD is used to describe the Right Ascension and combines
various names used over years and epochs e.g. RA(ICRS), RA1855, RAB1900, RAB1950, 
RAhms, RAJ2000 etc.
\end{itemize}

\section{UCD Topic Map}
\label{sect:ucdtm}  

A UCD Topic Map is constructed from the metadata (column names and units)
for a set of tables (catalogs). It does not concern itself with the
actual data in the catalog. Their sheer number can be overwhelming and 
that problem can anyway be tackled outside separately. Instead, the UCD Topic Map
revolves around Columns, Column Names and Tables. Thus,
the main topics include:
\begin{itemize}
\item all 1400+ UCD names, 
\item the corresponding UCD descriptions, 
\item the column Names from the Tables used in the Topic Map, and
\item the Tables themselves (any set of tables can be used to make a Topic Map).
\end{itemize}

In addition, there are several other topics owing to the hierarchical
nature of the UCDs (e.g. the depth of a UCD in the list), units associated
with different columns and any number
of external links per Table (e.g. a hyperlink to the actual table somewhere 
on the internet) or even per Column (e.g. a histogram for the column, or
quantiles obtained through external CGI programs). It is these external links 
that provide a great deal of flexibility to the Topic Map.

In the following few paragraphs we will be switching between descriptive and visual views
of different aspects of Topic Maps.
We show in \fig{} Fig.~1 the top view of the Topic Map as seen in Ontopia's 
freely available Omnigator\footnote{http://www.ontopia.net}. The view shows the 
different subject indexes (explicitly defined topics),
relationship indexes (defined associations), role indexes (the roles played
by the topics in the associations) and resource indexes (external resources
e.g. links to tables). As indicated before, the Topic Map is a network of
links allowing one to jump from a general topic (e.g. UCD) to a particular
instance of it (e.g. the UCD PHOT\_FLUX\_IR\_12) to the Table in which this UCD
occurs (e.g. the catalog of Seyfert Galaxies). The display with UCD as the
central topic is captured in \fig{} Fig.~2 while \fig{} Fig.~3 shows  
a view with PHOT\_FLUX\_IR\_12 being the central topic and \fig{} Fig.~4 
shows an HST related table to be the central topic. Each of these views
is rich in semantic information.
\begin{figure}
\begin{center}
\begin{tabular}{c}
\psfig{figure=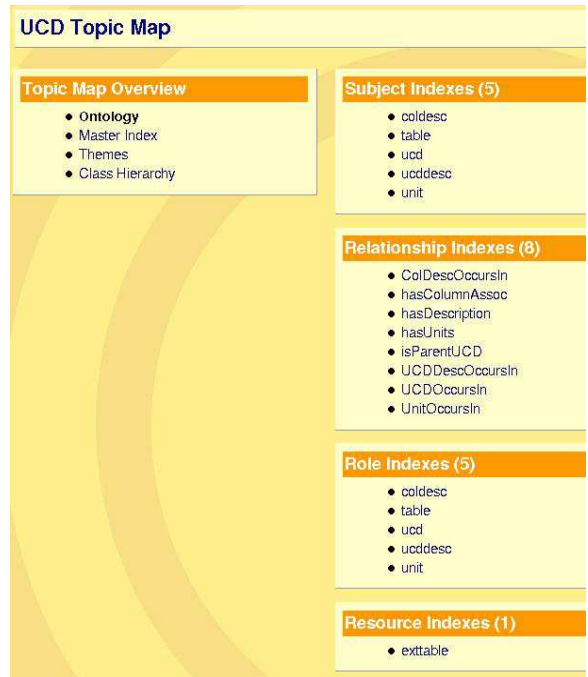,height=9cm}
\end{tabular}
\end{center}
\caption[figure 1] 
{ \label{fig:mainview} This is the top view of the UCD Topic Map. The main topics, associations,
roles played by topics in associations are seen. The main topics are like common nouns and they
denote various {\it topic types} that other topics have. Relationship Indexes are the different
associations that we have predefined. A user who is more interested in some other associationship
can easily get that added through the merging mechanism described in Sec.~5. The Role Indexes 
denote the roles played by the main topics in the associations. In our Topic Map the main topics do not play
multiple roles and hence the Subject Indexes and Role Indexes lists are identical. Finally, we have
illustrated the use of a single external resource viz. {\it exttable} i.e. a hyperlink to the actual table.
As elaborated in the text, this will actually be a long list of external links pointing to varied
VO resources including table merging, astrostatistics etc.}
\end{figure} 

For the UCD Topic Map we have defined very basic associationships between the different topics.
Some examples are:
\begin{itemize}
\item UCD {\it occurs in} Table,
\item UCD {\it has description} Column Description,
\item Column (of Table) {\it has units} Unit,
\item Column {\it occurs in} Table,
\item UCD {\it is parent of} UCD.
\end{itemize}
\begin{figure}
\begin{center}
\begin{tabular}{c}
\psfig{figure=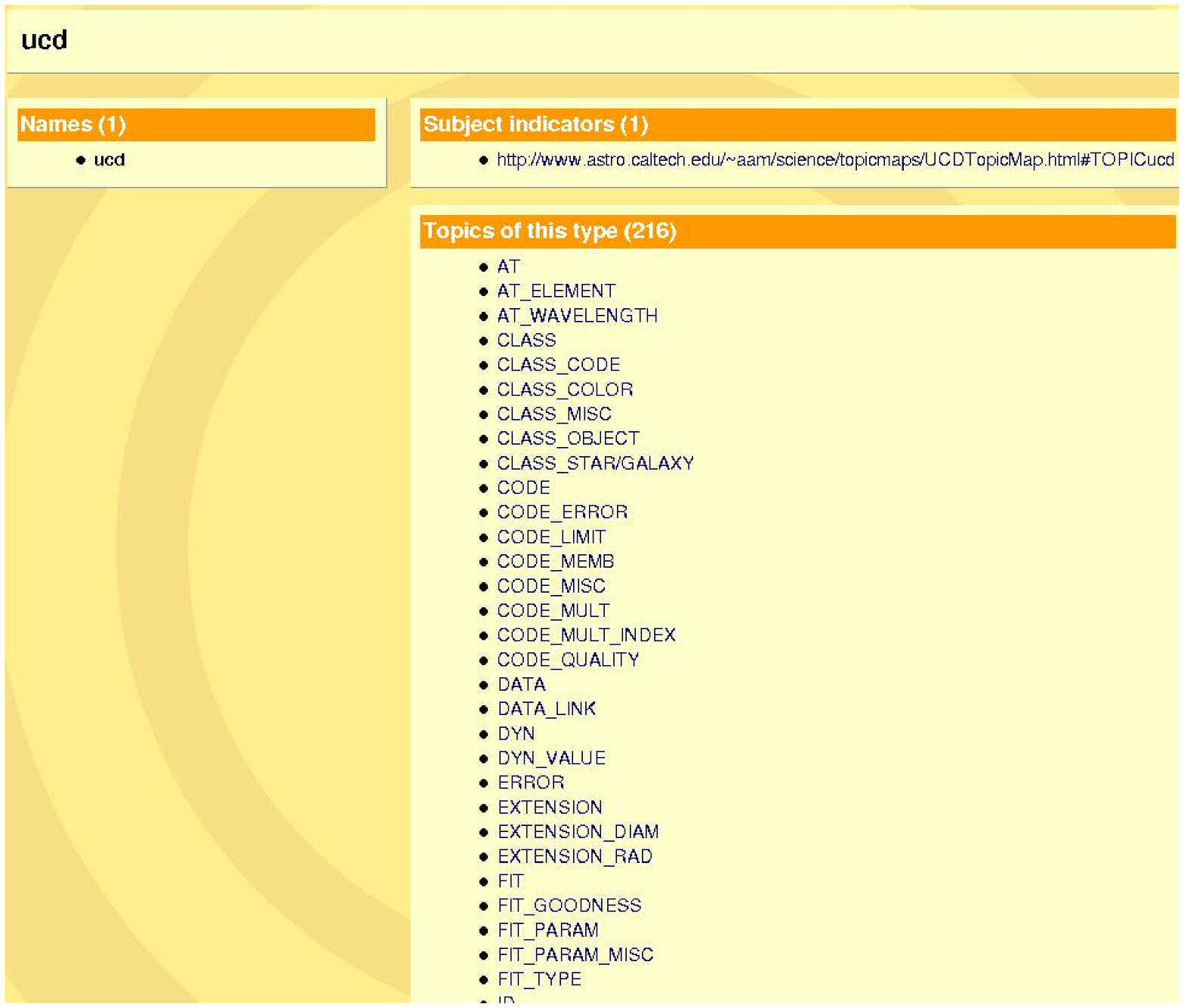,height=9cm}
\end{tabular}
\end{center}
\caption[figure 2] 
{ \label{fig:ucdlist} Topic Map view centered on the topic {\it UCD}. A partial list
of UCDs is visible. Each is a clickable hyperlink pointing to a separate page with
occurrences of that UCD. Selecting different UCDs will allow the user to explore different
regions of the data resource.}
\end{figure} 

Similarly, the occurrences in the UCD Topic Map can be grouped into a small
number of sets:
\begin{itemize}
\item UCDs occur in Tables,
\item Columns occur in Tables,
\item Units occur in Tables,
\item Tables occur as external links in many places.
\end{itemize}
\begin{figure}
\begin{center}
\begin{tabular}{c}
\psfig{figure=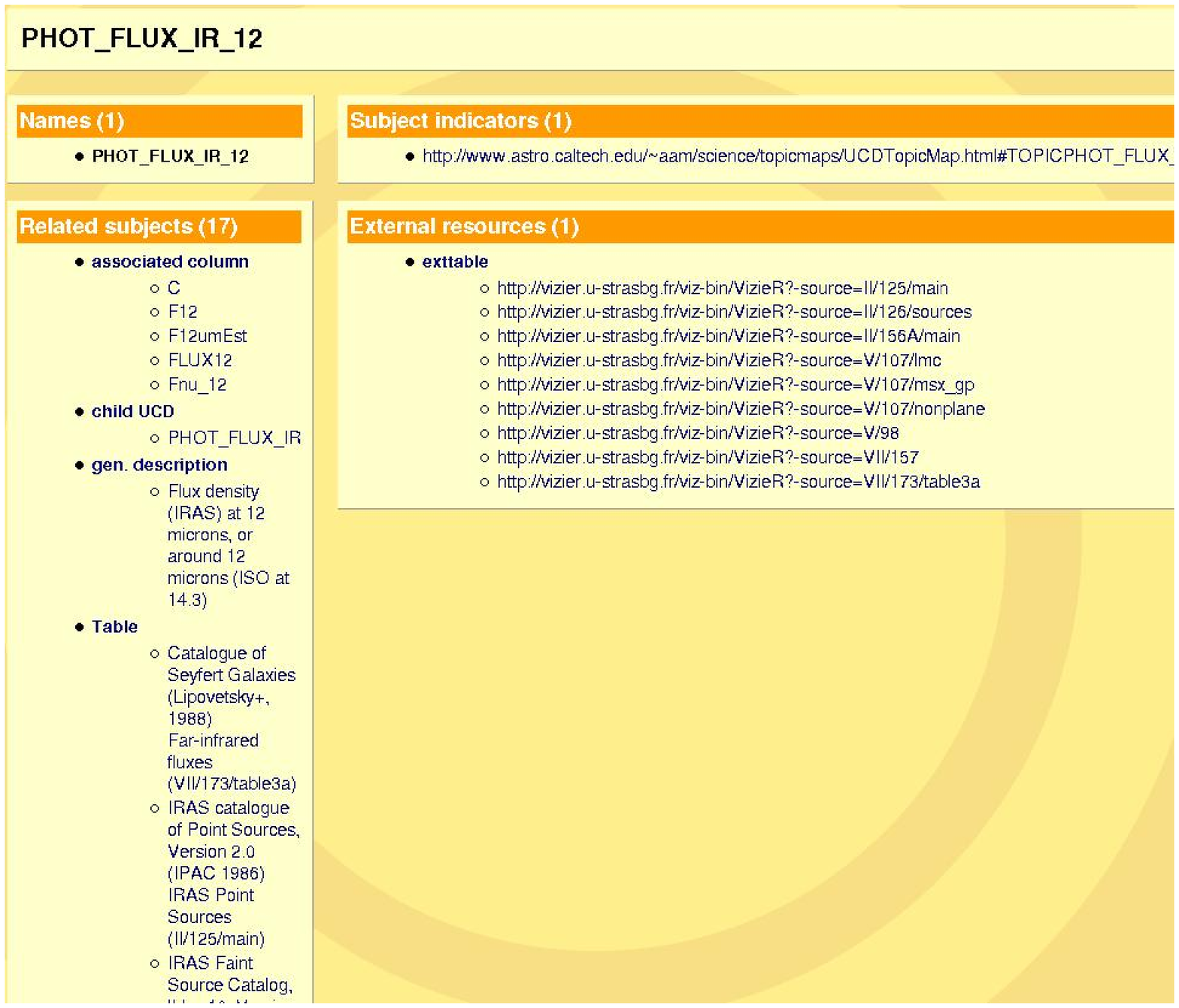,height=9cm}
\end{tabular}
\end{center}
\caption[figure 3] 
{ \label{fig:ucddescir} A separate page for the UCD PHOT\_FLUX\_IR\_12 is depicted here. With it are
seen the different tables it occurs in and the external links for those tables. Under ``associated column''
one can see the different names that different astronomers have used to denote the flux density at and
around 12 microns.}
\end{figure} 

If one is dealing with a single table, one encounters several columns, each column with
its corresponding UCD and a few of the columns having units. There are also several tens of
associations and occurrences. Together they form a structure
which is graspable without any sophisticated arrangements. The HST related view in \fig{} 
Fig.~4 is an example. It is when one goes to tens of catalogs that the complexity 
increases rapidly and a more advanced knowledge organization becomes necessary.
In the UCD Topic Map we present
here\footnote{http://www.astro.caltech.edu/$\sim\!$aam/science/topicmaps/ucd.html},
we have used the 100 most frequently accessed catalogs at Vizier.
A partial list of the catalogs is shown in \fig{} Fig.~5, and the 
resulting number of topics, associations etc. are partially sketched in \fig{} Fig.~6. 
It is with these large numbers that we can start asking interesting questions. This is where the
data/knowledge discovery begins.
\begin{figure}
\begin{center}
\begin{tabular}{c}
\psfig{figure=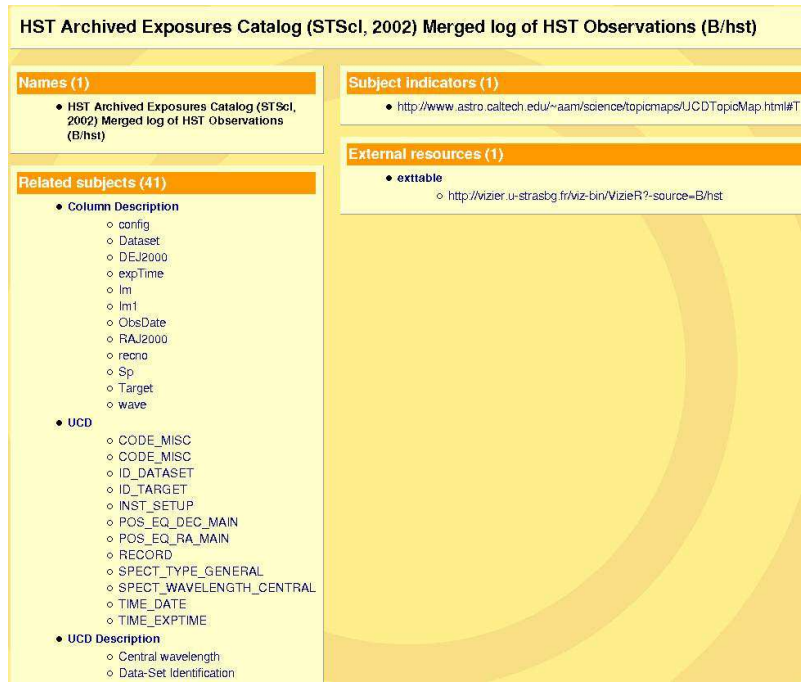,height=9cm}
\end{tabular}
\end{center}
\caption[figure 4] 
{ \label{fig:tabledesc} Typical page centered on a Table. The table ``HST archived exposures catalog'' 
is seen here.
Also seen are the different columns in the table, associated UCDs, and units when available.}
\end{figure} 
\begin{figure}
\begin{center}
\begin{tabular}{c}
\psfig{figure=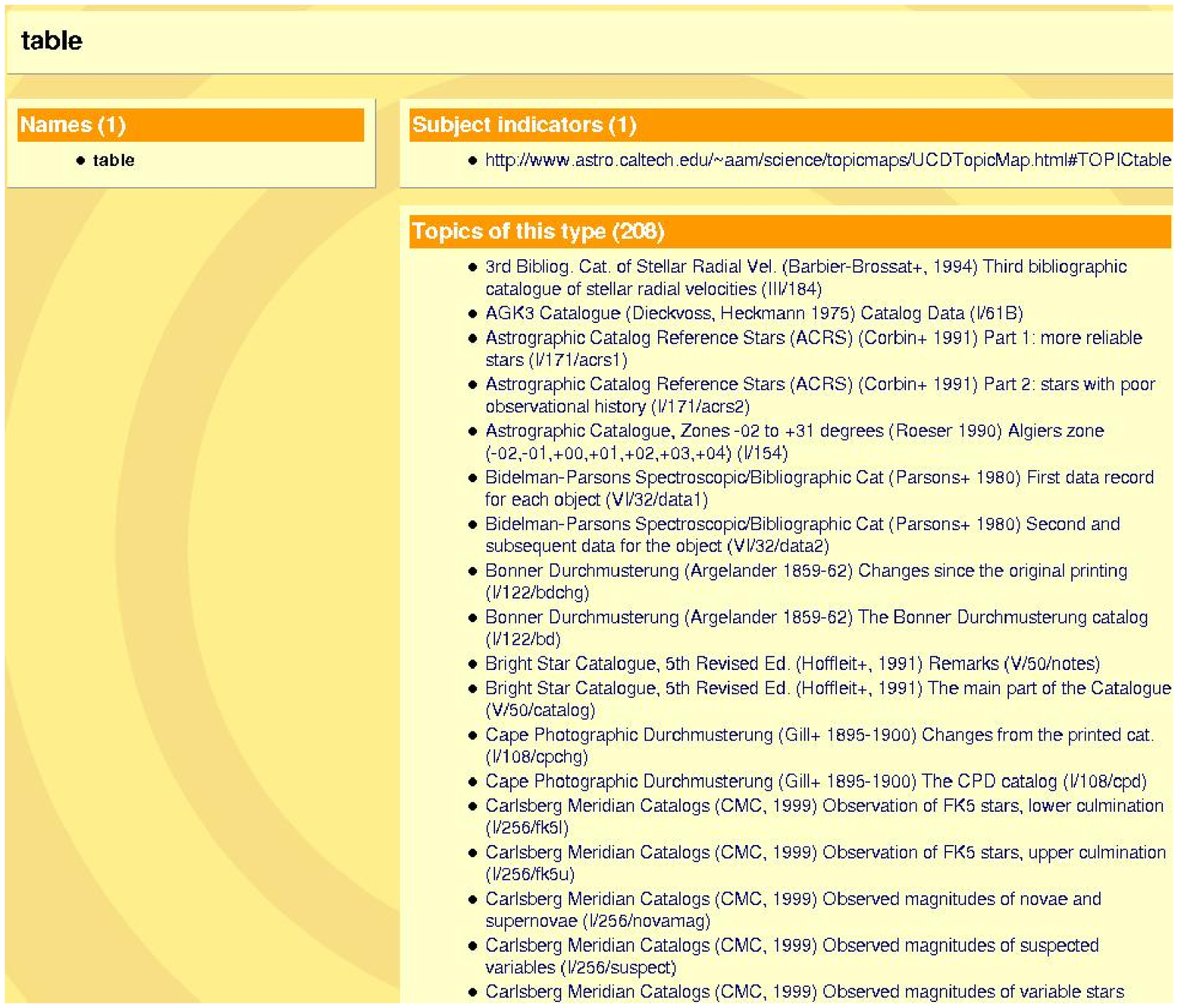,height=9cm}
\end{tabular}
\end{center}
\caption[figure 5] 
{ \label{fig:tablelist} The (partial) list of tables that constitute the current UCD Topic Map.
The individual tables can be separately browsed for columns, UCDs, units and external links.}
\end{figure} 
\begin{figure}
\begin{center}
\begin{tabular}{c}
\psfig{figure=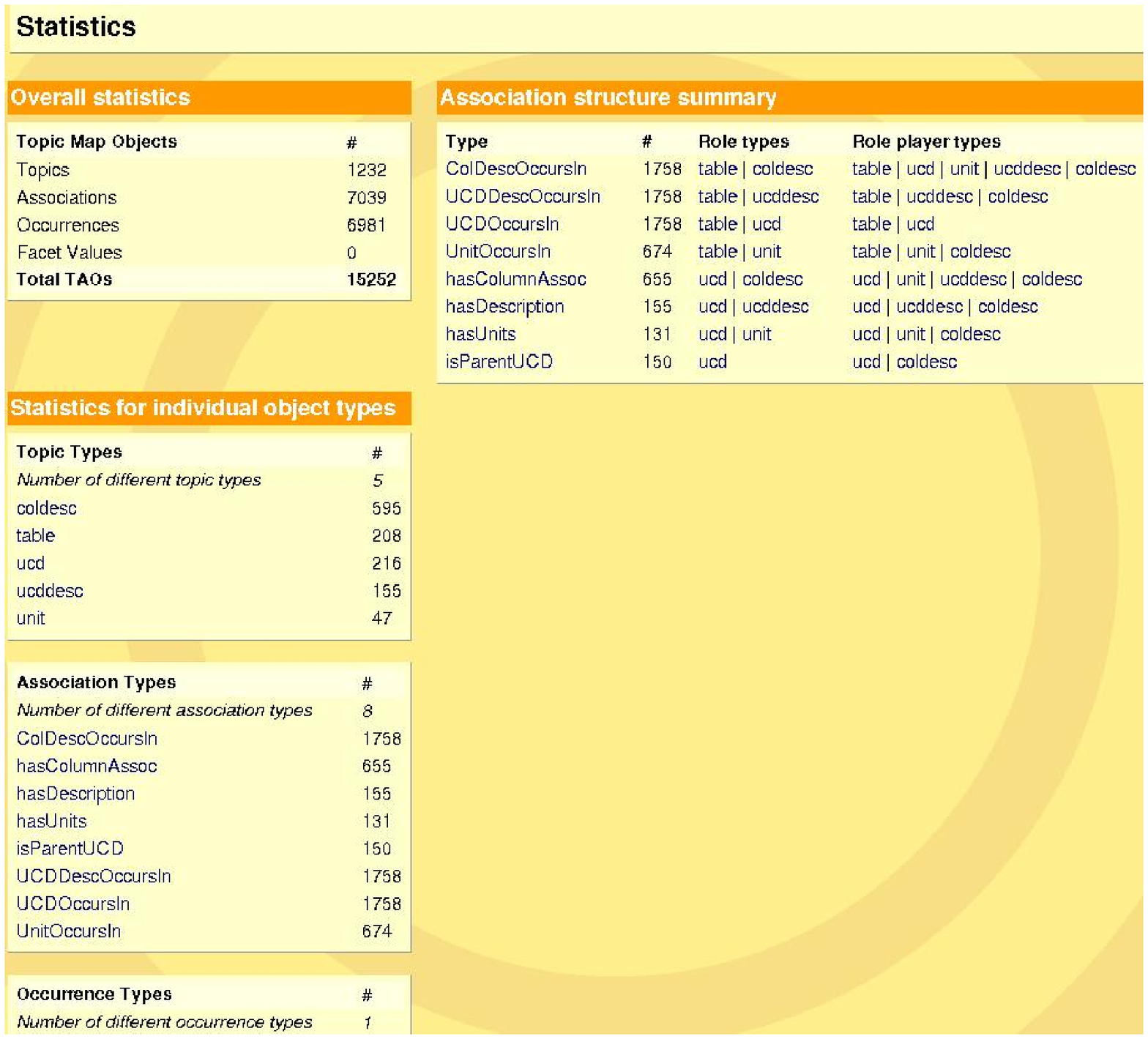,height=9cm}
\end{tabular}
\end{center}
\caption[figure 6] 
{ \label{fig:statsummary} A summary of topics, associations, occurrences etc. for the UCD
Topic Map made out of 100 most frequently accessed catalogs at Vizier. Thousands of topics,
associations and occurrences are a clear indication
that a good knowledge discovery tool is needed for exploring such a rich data resource.}
\end{figure} 

In case of a single table one has the table as the main topic with different columns present in the
table forming other topics, as also the UCDs associated with these columns and the Units
associated with some of the columns. A schematic of the scenario is presented in
\fig{} Fig.~7. When we move to a multitable UCD Topic Map, and select any UCD as the main 
topic, we can see all the associated tables (in which the UCD is present). Depending on one's 
interest, one can then start asking specific questions. One such scenario is summarized in
\fig{} Fig.~8. Consider an astronomer interested in the infrared part of the electromagnetic 
spectrum, in particular the 12$\mu$m flux density. She asks for tables
(within the 100 most accessed tables in the UCD TM)
that have a column which corresponds to the UCD PHOT\_FLUX\_IR\_12.
She is returned a list of such catalogs: Catalog of Seyfert Galaxies, IRAS catalog of Point Sources, the
extended $12\mu$m galaxy sample and so on. Each of these takes her to that table allowing her to explore
different columns present in those tables (e.g. which area has been covered, how faint they go, which
epoch it is, what the resolution is and so on). In addition, the Topic Map 
author can choose to add different scopes
to each of the table to, for instance, convey if the catalog is galactic or extra-galactic. Such 
information could then be used by the user
to determine if the catalogs contain columns on which a meaningful join can
be performed. A very VO like example would be: ``choose Elliptical galaxies that are strong in radio, fairly
bright in infra-red but do not contain an AGN''. By navigating through the appropriate keywords for
IR, radio fluxes and catalogs of elliptical galaxies and catalogs of non-agn galaxies such a search could
be achieved. Additionally the Galactic catalogs need not all be thrown away. They can be used to check
if there exists a population of objects that can act as a contaminant to the population that the user
has chosen.
\begin{figure}
\begin{center}
\begin{tabular}{c}
\psfig{figure=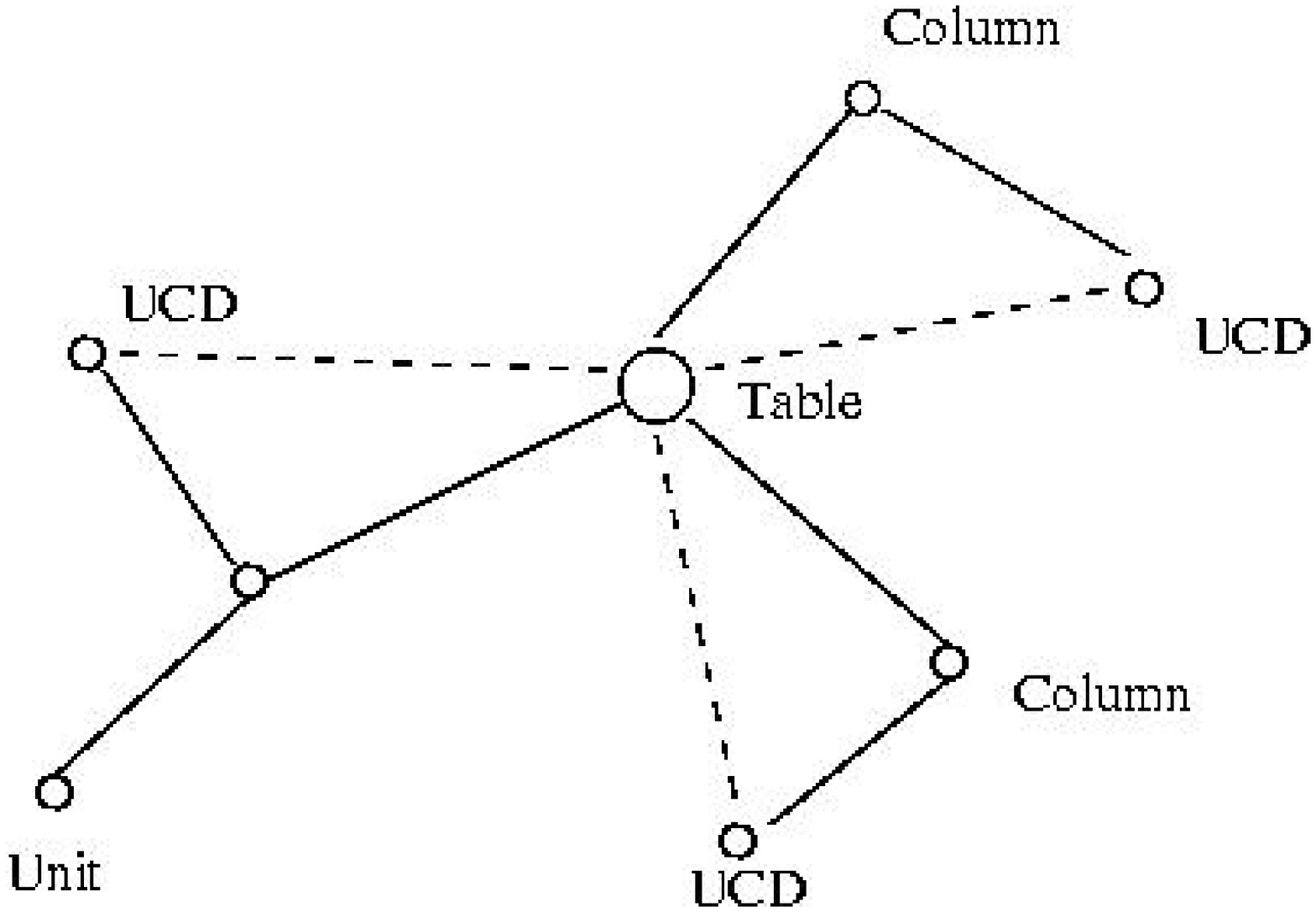,height=9cm}
\end{tabular}
\end{center}
\caption[figure 7] 
{ \label{fig:singletable} A single table has just a few columns, UCDs and even fewer units and is easily
navigable.  }
\end{figure} 
\begin{figure}
\begin{center}
\begin{tabular}{c}
\psfig{figure=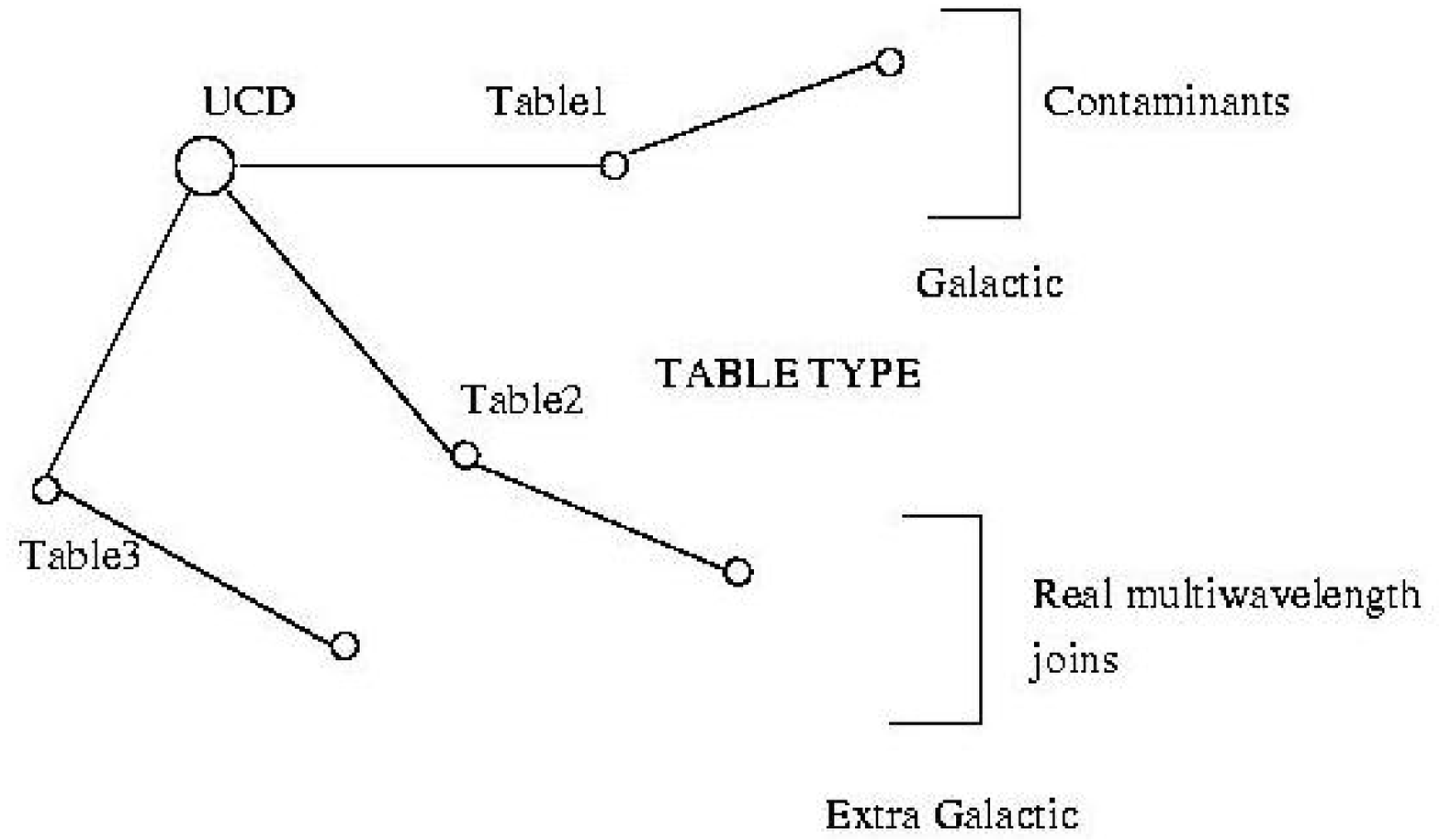,height=9cm}
\end{tabular}
\end{center}
\caption[figure 8] 
{ \label{fig:multitable} When multiple tables are put together, the complexity increases manifold.
With Topic Maps the complexity can be harvested to ask interesting questions and get meaningful answers.}
\end{figure} 

Thus data discovery is all about asking the right questions: Can I merge X-Ray and Radio
catalogs? Which ones? Do their units match? What are the parameter ranges? What is the basic statistics
for them? How does the histogram for a column look and so on.
Some of the possibilities mentioned above can also be carried out using tools available 
elsewhere (e.g. at CDS).
However, segregation of different entities by the scoping mechanism affords Topic Maps a stronger
semantics allowing a quicker zooming in on the required data. Another aspect is the ability to merge 
different Topic Maps. We discuss it in the next section.

\section{TOPIC MAP MERGING}
\label{sect:tmmerge}  

Topic Map merging is not the same as merging of columns from different tables to create a database join.
Topic Map merging involves identifying identical Topic Map {\it topics} in different Topic Maps and merging
their associations and occurrences to give rise to a single merged topic. For doing this, a
{\it SubjectIdentity} tag is used. As a result, one can have two topics with the same name if they have
different SubjectIdentity. This is in fact desirable since we often use the same name in different
contexts for different entities e.g. {\it satellite} could be a general term for the satellite of a planet,
or it could be a man-made communications satellite.

Here we are more interested in topics that can be merged. Continuing to discuss our hypothetical astronomer,
now she is equipped with her own dataset of X-Ray observations which she has put in the standard XML
format after having identified proper UCDs for the different column names. She would now like to merge
the database with the readymade UCD Topic Map (of the 100 most used catalogs) in order to later look
for the non-AGN radio ellipticals which are also detected in X-Ray. The Topic Map generator that
we have built allows precisely this to be done. It takes as input a list of tables and then generates a 
Topic Map out of it in exactly the same format as the UCD Topic Map (the new Topic Map is also a UCD Topic
Map, though custom built). The topic map then appears as a new entry in the list of Topic Maps
available. The {\it merge} button available at the top prompts the user 
to choose which Topic Maps are to be merged
(see \fig{} Fig.~9). 
Once the user chooses the original UCD Topic Map and the new custom built TM, the merging is done and a new 
{\it merged} topic map is available for exploring. The cgi-bin form through which the topic map
generation is carried out is shown in \fig{} Fig.~10.  A flowchart of the merging procedure is  
shown in \fig{} Fig.~11. 
The TM generator can also make a Topic Map from just the list of catalog names in the format Vizier uses.
It downloads the catalogs and any sub-catalogs (currently only from Vizier) and forms the Topic Map using
the template built
into the program.

\section{QUERYING AND INDEXING}
\label{sect:queryI}  

Lastly we talk about two aspects that make knowledge discovery with Topic Maps even more
viable. These are querying and indexing. We mentioned earlier that Topic Maps are XML structures
working on other XML structures and that querying XML is not straightforward in the usual way.
The ISO community has not formalized a TMQL (Topic Map Query Language) yet. However, many precursors
like eTMQL\footnote{empolis: http://k42.empolis.co.uk/tmql.html}, AsTMa\footnote{Bond University,
Australia: http://topicmaps.bond.edu.au/astma} and Tolog\footnote{Ontopia: http://www.ontopia.net} exist.
We found Tolog easy to use and it comes bundled with Ontopia's free Omnigator.
The syntax used is a combination 
of an SQL like language and the powerful prolog. The use of prolog's power allows one to make
extensive searches. Examples of some queries are given below.
\begin{itemize}
\item query: List all UCDs \\
 code: instance-of(\$A,ucd)? \\
 meaning: give me all instances that match \$A=ucd (\$A is a variable). 

\item query: Count number of child UCDs \\
 code: Select \$A, count(\$B) from isParentUCD(\$A : ucd, \$B : ucd)? \\
 meaning: Count all \$B such that \$A and \$B are both UCDs and \$B is the child of \$A. 

\item query: Order the above in descending order, sorted on count. \\
 code: Select \$A, count(\$B) from isParentUCD(\$A: ucd, \$B: ucd) order by \$B desc? \\
 meaning: Select \$A such that \$A and \$B are UCDs, count the number of \$Bs and put
it in inverse numeric order. 
\end{itemize}

The queries can be executed on individual Topic Maps or on merged Topic Maps. Unlike an SQL
query run on a database, all the returned entries are hyperlinks too, allowing the user to
jump to any of the results and continue the data discovery trails.

Indexing is the other important aspect. Topic Maps began life with indexing in mind. It is
not surprising that it continues to be one of the strengths of this tool. One can create different 
indexes per Topic Map so that searching is faster.  As for queries, searches return hyperlinked documents
and also the type of match it found and the confidence level of the match. The hyperlinked searchable
multiple indexes make a Topic Map a combination of indexes, thesauri and glossaries.
\begin{figure}
\begin{center}
\begin{tabular}{c}
\psfig{figure=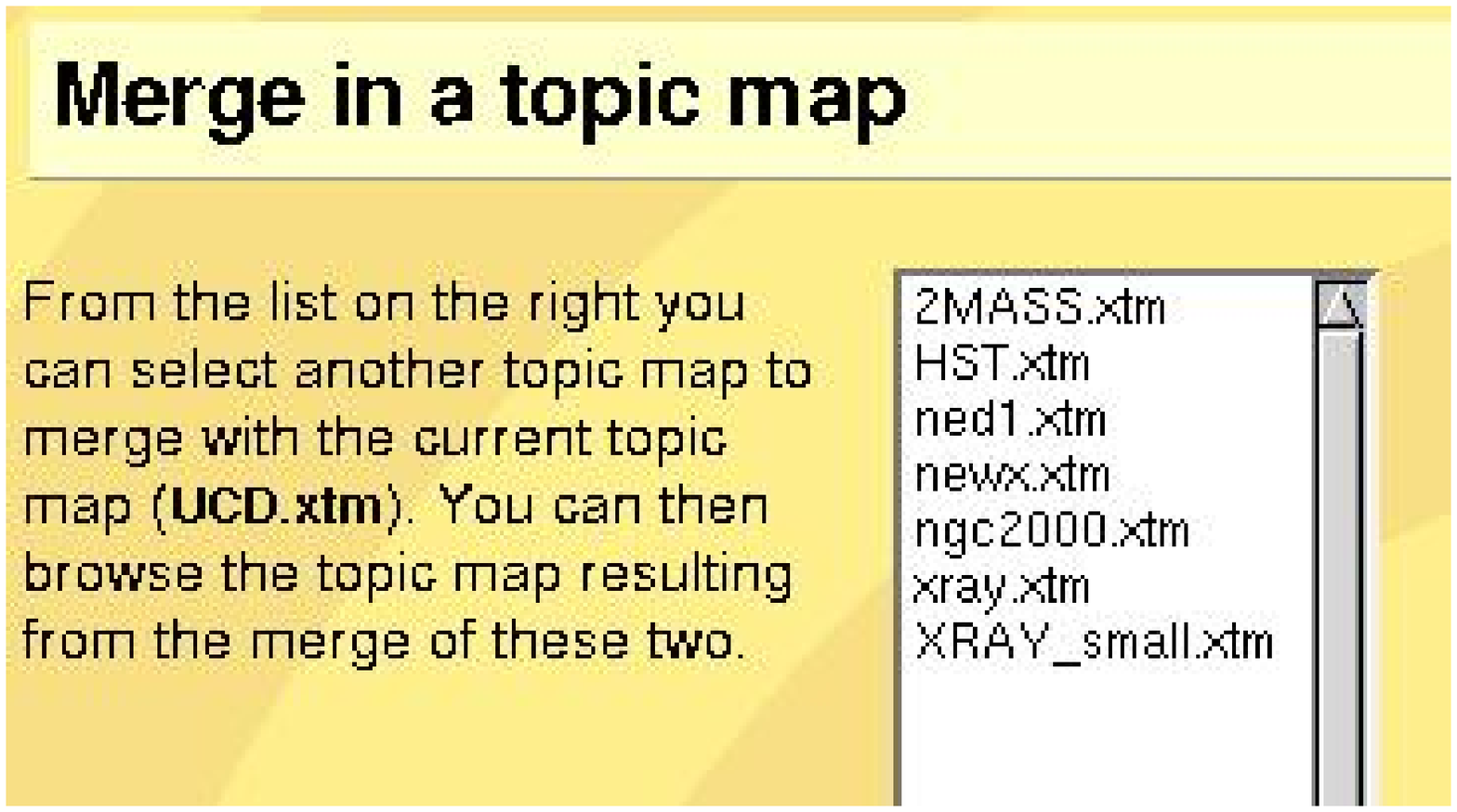,height=8cm}
\end{tabular}
\end{center}
\caption[figure 9] 
{ \label{fig:tmmerge} The form that allows users to merge different Topic Maps. Associations and occurrences
of Topics which have identical {\it SubjectIndicators} are merged.}
\end{figure} 
\begin{figure}
\begin{center}
\begin{tabular}{c}
\psfig{figure=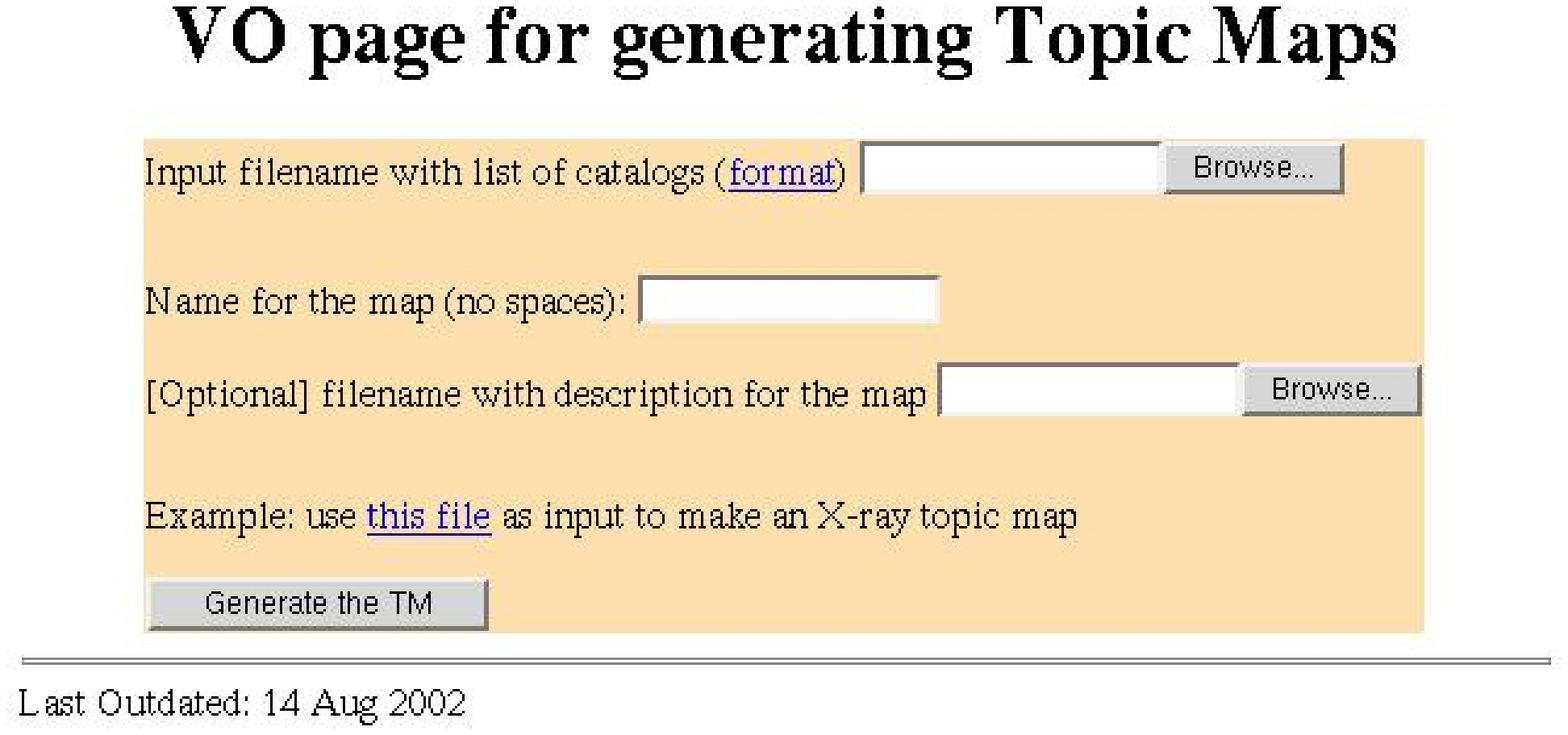,height=9cm}
\end{tabular}
\end{center}
\caption[figure 10] 
{ \label{fig:tmgenerator} The cgi-bin form that allows users to generate their own UCD Topic Maps
from confirming tables (XML format). Such a Topic Map can then be merged with other existing UCD Topic Maps.}
\end{figure} 
\begin{figure}
\begin{center}
\begin{tabular}{c}
\psfig{figure=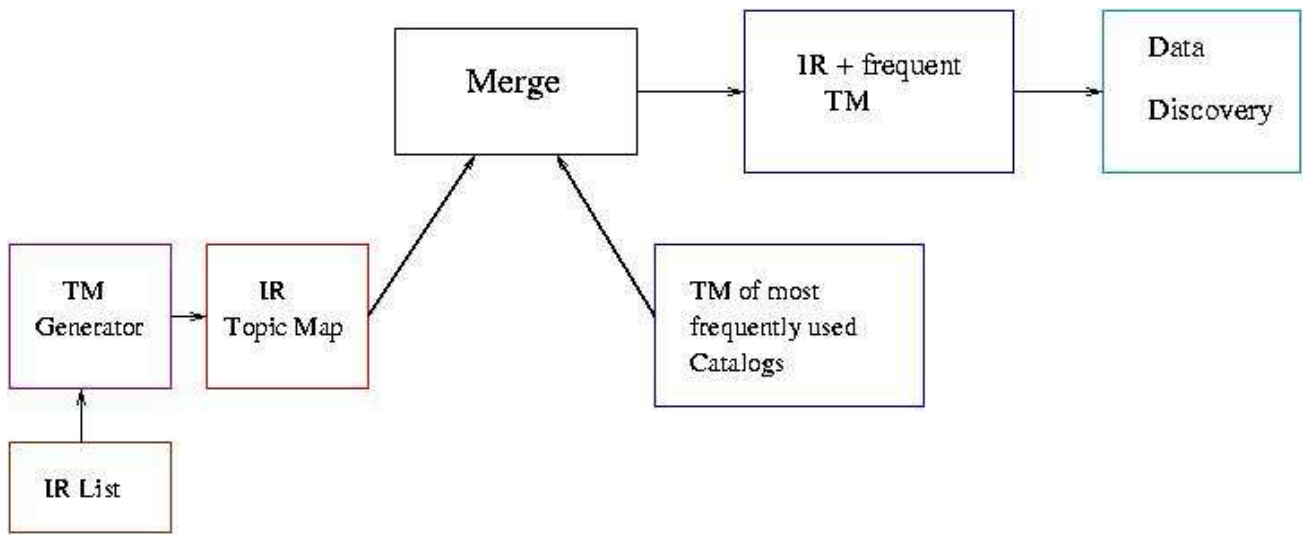,height=7cm}
\end{tabular}
\end{center}
\caption[figure 11] 
{ \label{fig:mergeschem} A flowchart showing the path from individual datasets to data discovery through
the medium of Topic Maps and their merging.}
\end{figure} 

\setcounter{footnote}{1}
\section{DISCUSSION AND CONCLUSIONS}
\label{sect:disc}  

We have outlined  above the construction and use of {\it a} UCD Topic Map and how users
can create their own UCD Topic Maps and merge them with other UCD Topic Maps. While this can
already act as a tool for custom viewing of data and as a knowledge discovery tool, much
needs to be accomplished. We will keep adding bells and whistles to the Topic Map
page\footnote{http://www.astro.caltech.edu/$\sim\!$aam/science/topicmaps}. 

A lot of stress has been placed on the external links. This is because Topic Maps need not reinvent wheels
for actions that other services are capable of e.g. Topic Maps facilitate selection of columns for merging.
One
could then easily use the merging VO service provided by
Roy Williams\footnote{http://virtualsky.org/conesearch-xmatch.htm} to do the actual merging.
Similarly, for statistics of columns, one could use the prototype
astrostatistics\footnote{http://avyakta.caltech.edu:8080/cgi-bin/astrostat3.cgi} service.
In passing, it must be stressed that Topic Maps are a general knowledge organization
tool and many varied Topic Maps can be constructed with the VO in mind. Some example sketches regarding
literature searches and observing logs were presented in \REF Mahabal et al.(2001).

We encourage you to explore Topic Maps, generate some of your own, UCD or otherwise. Your comments are most 
welcome.
\acknowledgments     
This work was supported in part by
the NASA AISRP program and the NSF ITR program.
Many thanks to Vidyullata Mahabal for help with the figures.
Many thanks to Francois Ochsenbein for patiently replying queries about UCDs.




\end{document}